**Online Assemblies: Civic Technologies Reshaping the Public Space**

Tallulah Frappier

Speaking or writing of political assemblies tends to evoke the action of people gathering to deliberate, or the spaces in which this deliberation might take place. One thing that is often overlooked, however, is the fact that these spaces can be digital. Online assemblies have become more widespread in recent years; from the first Web forums to civic technologies specifically designed to host collective political debates. As digital services affect our possibilities for political mobilization and participation, I will here attempt to define the qualities specific to online assemblies, and to identify several patterns and continuities in the design features of civic technologies offering online spaces for debate.

**The Spread of Online Assemblies**

Since the creation of the Internet, digital spaces have been considered to be fruitful grounds for the creation of assemblies. In 1962, Doug Engelbart, then the director of Stanford's Augmentation Research Center, presented several uses for the computer. This demonstration, subsequently named "The Mother of All Demos,"[1] insisted on the possibility of computers becoming tools for collaboration and enhancing collective intelligence. Digital pioneers often held the belief that technology would help them change society without needing to acquiesce to existing power structures, or to devise and enforce new ones. As digital pioneer John Perry Barlow stated in his 1996 "Declaration of the Independence of Cyberspace": "We are creating a world where anyone, anywhere may express his or her beliefs, no matter how singular, without fear of being coerced into silence or conformity."[2] If these claims seem a little simplistic or indeed even naive today, they also help us to understand how digital tools have long-since been both imbued with and invested in the capacity to create new vehicles for collectively discussing ideas.

In 1985, *The Well*, one of the first and most famous online communities, was created. It was hosted by a specific form of discussion site: the Internet forum. Originally inspired by bulletin boards, these sites differ from discussion chats as their structure is asynchronous. Forums spread online under slight variations, enabling individuals and communities gather around particular topics. I decided here to identify them as assembly spaces due to the nature of their user-audience, their component functions, and their architecture. In simple terms, these sites aim to gather a large group of people together in order to discuss specific topics, within dedicated areas. Sometimes these discussions also result in a voting process, in which pre-posted messages can be voted on by other users, then ranked to reorganize content so the most "relevant" messages (as determined by this voting) appear first.

The central argument of "A Declaration of the Independence of the Cyberspace"—that the Internet enables the creation of an alternative space that cannot be governed by any outside force—has been rapidly eroding since the late 1980s. Through observing increasing interferences in the Internet ecosystem led by both the state and corporations alike, digital activists have grasped the intertwining of cyberspace with civil society, leading them to instead invest their time and energy in fields such as education, or political representation.[3] Creating alternative, liberated democratic spaces on the Internet was never going to be possible if the existing order of the offline world was not also simultaneously changed. Therefore the goal of digital activism is currently more closely focused on creating ways to complete, fix, or challenge an overall political macrocosm. It is with this aim in mind that "civic technologies" have emerged. This term embraces various kinds of initiatives that use digital resources to change political rules, or to intensify political engagement.[4] They attest to an increasing crisis of confidence in representative democracy, and revolve around the place of digital technologies—both in institutions and counter-power organizations. Academic and non-academic literature has referred to civic technologies from both government-centric and citizen-centric perspectives.[5] When civic technologies are government-centric, they "enable top-down change through the promotion of government transparency, accessibility of government data and services and promotion of civic involvement in the democratic process,"[6] and they can also be described as citizen-centric; as "platforms and applications that enable citizens to connect and collaborate with each other and/or with the government."[7] Online collective deliberation therefore constitutes a significant challenge in government-centric and citizen-centric initiatives, and many civic technologies offer services for people to debate together about specific issues, whether in public or in private contexts.

---

[1] Doug Engelbart, presentation at the Fall Joint Computer Conference on December 9, 1968, San Francisco. See: https://www.dougengelbart.org/content/view/374/464/#2.

[2] John Perry Barlow, "A Declaration of the Independence of Cyberspace," published online on February 8, 1996. See: https://www.eff.org/cyberspace-independence.

However, these technologies' purpose is not to promote specific democratic subjects, but to intensify democracy through their functionalities. They induce a procedural vision of politics in which good procedures allow for good democracy, driven by the central question: "How do we make democracy work?" Examining the structure of these online assemblies is therefore fundamental to understanding the democratic conceptions they have come to materialize.

**An Overturn of Assemblies and Deliberation**

Online, every moment of action and interaction is authorized and delimited by technical resources. This phenomenon led Lawrence Lessig to famously state that "Code is Law,"[8] arguing that architecture—taken here to mean the physical traits and qualities of any given space—is the primary source of regulatory principles online. This assertion has since been qualified,[9] but it remains the case that technical choices and determinant necessities online can drastically affect the "canonical" shape of the assembly and of one of its intrinsic social practices: deliberation.

---

[3] Nicolas Auray and Samira Ouardi, "Numérique et émancipation. De la politique du code au renouvellement des élites," *Mouvements*, vol. 79, no. 3 (2014): 13–27.

[4] Dominique Cardon, *Culture numérique* (Paris: Presses de Sciences Po, 2019), 277.

[5] Jorge Saldivar, et al., "Civic Technology for Social Innovation," *Computer Supported Cooperative Work* 28, no. 1 (2019): 169–207.

[6] The Knight Foundation, *The Emergence of Civic Tech: Investments in a Growing Field*, (February 26, 2014), https://knightfoundation.org.

[7] Manik Suri, "From Crowd-sourcing Potholes to Community Policing: Applying Interoperability Theory to Analyze the Expansion of 'Open311'," *Berkman Center Research Publication* 2013, no. 18 (August 2013): 1.

[8] Lawrence Lessig, *Code: And Other Laws of Cyberspace* (Sidney: ReadHowYouWant.com, 2009).

[9] Françoise Massit-Folléa, Cécile Méadel, and Laurence Monnoyer-Smith, eds., *Normative Experiences in Internet Politics* (Paris: Presses des Mines, 2012).

If we return for a moment to the notion of the "forum"—its name, as is well known, directly evokes the Latin *forum*, a public place for commercial and political exchanges. However the term itself is more likely drawn from the conception of the traditional public forum in the United States Constitutional Law, in which a "forum" is understood to be a public place dedicated to assemblies and debate. It's interesting to note that, while being differentiated from the synchronous chat, the forum has drawn its name from a space reliant on a common temporality. The most significant aspect of the forum highlighted online is therefore its spatial quality, its interlinking functions as they evoke another term: "synchorization." Synchorization is the spatial counterpart of synchronization, and has been described by Boris Beaude as a "process consisting in having a common place for being and doing,"[10] that allows people to interact.

This asynchronous aspect of the assembly induces a drastic change of the traditional conception of debate. Debating is no longer a purely back-and-forth, responsive practice. Online, one has to wait to read a counter-argument. The online forum debate is constructed over a long period of time—over days, weeks, or even years. This time dilation can function as a mechanism encouraging constructive dialogue, but it can also make communication harder.[11] Helpful protocols and procedures like speaking turns or time management no longer hold relevance, as everyone can express themselves at any time. If a selection of content has to be made it happens after debate, and is qualitatively assessed according to questions like: "Is it relevant to the subject?" or "Has it already been articulated?" Moderation, however, can take place either before or after the publication of a post—whether this is channeled through the decision of an administrator, or through reports from the community.

Synchorization is therefore enabled by online representation—you cannot be "physically" online, but rather you are represented by avatars, names, visualizations, and the words you write. Deliberation is no longer about speech, but about written practice. To analyze parliamentary assemblies' architectures, French political scientist Jean-Philippe Heurtin highlighted the fact that the multiple verbal and nonverbal interactions that take place between points of views, bodies, and facial expressions have shaped the course of interpersonal debate.[12] Online, the interactions of these key elements disappear. One could perhaps argue that this change is for the best: it could help debate to become more rational, and redirect the focus on the content of the argumentation itself. Sometimes, it can even lead to "benign disinhibition,"[13] where people are more inclined to be kind to people they can't physically see. But it can also complicate the identification of a common ground,[14] and boost cyberbullying,[15] trolling,[16] and incivility.[17] Heurtin also evoked a "theater of glances" in order to describe amphitheater-shaped assemblies as glance-organizing machines that frontally oppose the spectator and the one speaking. Online assemblies constitute an opposite model. Within an online debate, internauts are deprived of any specific points of view, as they all see the same interface. It's only when they submit a form of writing (or act through voting) that they identify themselves as being involved in a debate. Beyond these particular forms of action, they can stay "pure spectators,"[18] which is to say, should they wish to they can remain fully disengaged.

---

[10] Boris Beaude, *Internet: changer l'espace, changer la société* (Paris: Éditions FYP, 2012).

[11] Amanda Baughan, et al., "Someone Is Wrong on the Internet: Having Hard Conversations in Online Spaces," *Proceedings of the ACM on Human-Computer Interaction* 5, no. CSCW1 (April 2021): 1–22.

[12] Jean-Philippe Heurtin, *L'Espace public parlementaire: essai sur les raisons du législateur* (Paris: Presses Universitaires de France, 1999).

[13] John Suler, "The Online Disinhibition Effect," *Cyberpsychology & Behavior: The Impact of the Internet, Multimedia and Virtual Reality on Behavior and Society* 7, no. 3 (July 2004): 321–6.

Heurtin also described an absence of endodeterminism within parliamentary space: the universal audience is outside of the assembly, and people who debate have to "talk to the windows."' Online, the universal audience is potentially within the assembly, but we still have to talk to it through windows, albeit browser windows. Online assemblies are not closed spaces. Their openness is underlined by their structure of hyperlinks. Discourses might also consist of the aggregation of copy-paste texts, and almost every issue can be shared on other platforms and communities thanks to dedicated icons.

|       |                  | Offline Assemblies       | Online Assemblies       |
|-------|------------------|--------------------------|-------------------------|
| Where | Place            | Local                    | Distributed             |
| When  | Synchronicity    | Synchrone                | Asynchrone              |
|       | Time continuum   | Time density             | Time dilation           |
| Who   | Scale            | A sample of people       | Virtually everybody     |
| How   | Type of practice | Speech practice          | Written practice        |
|       | Point of view    | Specific                 | Not specific            |
|       | Representation   | Bodies                   | Images                  |
|       | Identify as an actor | By being in the assembly | By acting in the assembly |

**Fig.2** Comparative table of online and offline assemblies' specificities.

---

[14] Maciek Lipinski-Harten and Romin W. Tafarodi, "Attitude Moderation: A Comparison of Online Chat and Face-to-face Conversation," *Computers in Human Behavior* 29, no. 6 (2013): 2490–2493.

[15] Hinduja, Sameer, and Justin W. Patchin. "Cyberbullying: An exploratory analysis of factors related to offending and victimization." *Deviant behavior* 29.2 (2008): 129–156.

[16] Justin Cheng, et al., "Anyone Can Become a Troll: Causes of Trolling Behavior in Online Discussions," *Proceedings of the 2017 ACM Conference on Computer Supported Cooperative Work and Social Computing* (February 2017): 1217–1230

[17] Porismita Borah, "Does It Matter Where You Read the News Story? Interaction of Incivility and News Frames in the Political Blogosphere," *Communication Research* 41, no. 6 (2014): 809–827.

[18] Jean-Philippe Heurtin, "Architectures morales de l'Assemblée Nationale," *Politix* 7, no. 26 (1994): 109–140.

These assemblies also showcase their assemblage nature. Each space is a knot of technical layers, of diverse functionalities, of people, organizations, links, topics, sources, etc. As for the debate itself, since it is asynchronous, everyone participating is already immersed within a completely different physical background, with varying affects on each user. As time goes by, debate winds into an increasingly tight knot of circumstances—one person will read a part of the debate, but never all of it, and will never be connected to another user at a specific time without ever reading the following argument. The act of deliberation is therefore obviously intertwined with outside places and moments.

**Design patterns**

As I touched on earlier, these "civic technologies" gather together various kinds of initiatives that use digital resources to change political rules, or to intensify political engagement. Directories of civic technologies can be found online and can help users identify which civic tools are proposing online assemblies as services.[19] Analyzing their interfaces can also help determine patterns in their designs. Though they all include debate features, these technologies are diverse. Some can be used in a specific timeline and with a specific community, instantiated by public or private authority (Assembl, Consul, Decidim)—while others are accessible to all and are not meant to deliver results to any particular sponsor (DebateArt, DebateIsland, or Communecter). This could indicate that we will face a wide range of debate features' designs.

First, the online forum appears to be a blueprint for their debate tools, presenting as they do the same core characteristics: they are asynchronous; they are organized in threads, posts and comments; and they present the same specificities that bear the potential to overturn the necessary conditions for deliberation.

Each of these selected debate interfaces follow roughly the same pattern of display. The top of the page is framed by an area dedicated to the topic containing a title, a small blurb, and sometimes a visual accompaniment. This is followed by an interactive space, offering users the option to vote either "for" or "against" the topic at hand, and typically displaying a running percentage of votes so far received. Sometimes this is just a posting area, with space to elaborate a small commentary on whether the user is for or against the topic, which might result in deliberation being part of a voting process. The lower part of the interface displays the deliberative space and its arguments. This conception seems somewhat counterintuitive when considered in comparison to typical processes of deliberation, in which voting takes place *after* a process of deliberation. Moreover, as the voting results are most of the time displayed, you see them even before reading the very core of the topic.

However, some of these platforms do not follow this form, such as DebateArt, or debate.org. Similar in form to some of the aforementioned oldest forms of cyberspace activism, these sites aim to actively foster deliberation between people, and are not linked to any institutional decision-making process. Each topic leads to two people arguing on opposing sides, and passes through several rounds. External users can then comment and vote for particular arguments, before a "winner" is finally declared.

---

[19] See: https://civictech.guide; http://civictechno.fr/civic-theque; and *Démocratie ouverte*, which gathers "democratic innovators," https://www.democratieouverte.org.

Within other interfaces, placing the voting process ahead of the deliberation space seems strange: why debate if your decision has already been made? These interfaces demonstrate multiple voting functionalities—from the vote on the main proposition, to votes on arguments and comments. This prioritising of the vote can also be crucial for articulating and promoting debate results and fostering participation from users, particularly when the given technology is supported by sponsors who require visible results. Voting results are an easier form of data to count and to reorganize than arguments are—they can be understood quickly and can facilitate swift reporting and decision-making processes, though of course they lack the subtlety that analysis of a range of arguments can generate. Furthermore, as numerical functionalities, they assimilate very well with digital environments and can be easily implemented. However, civic technologies that are not instantiated by sponsors but rather are animated by communities of citizens also present a profusion of voting functionalities. This could be prompted by a real need from the users, perhaps to have an overview of opinions on a debate—or by the fact that voting functionalities are typically constituted by a pattern instigated by the designers. Nevertheless this prioritization of voting functionalities often leads these forms of assembly to appear more like survey tools than deliberative spaces. I must note that some of them tend to minimize the significance of voting, by making voting buttons smaller or less central than other platforms, or by sequencing deliberation into phases of debate and phases of vote.

Two major design patterns can be identified in the display of arguments: the first gathers platforms resembling traditional Internet forums, whereby deliberative arguments form "posts" appearing as comments within threads. A second pattern is a poll-like, two-column feature in which users choose between a "pros" and a "cons" column before posting their arguments. Some platforms combine these two patterns—with arguments placed in threads, and tagged as either pros or cons throughout. Each of these patterns reveals different concepts of deliberation, and emulates various democratic practices. The first forum-inspired design involves an approach where participants can comment on an issue, but the common construction is challenging. The second, poll-like pattern polarizes the debate between the pros and cons and maintains the deliberation on a pre-established binary scale.

The second pattern (and the overall presence of votes) is also linked to arguments that online assemblies bear the potential to facilitate the exacerbation of opposing views. This potential for increased antagonism is fostered on two levels: first through the framework establishing that an argument has to be registered as either for or against an initial affirmation or question. Second, through the presence of voting functionalities attributed to questions or comments, usually signified by a like or an unlike, a plus or a minus. This seems to give legitimacy to an argument following an aggregative logic: the more people agree with it, the more so-called relevance it accrues. Certain configurations slightly differ from this binary positioning, such as Assembl—which follows the same poll-like pattern, but also bears a third column dedicated to alternative propositions—or Kialo, which enables space for more nuanced reactions to an argument's quality. However, online debate predominantly takes form through the opposition of arguments, and tends to be symbolically resolved through a vote between opposing sides.

These civic technologies thus display a very limited range of design patterns. There are several possible explanations for their aesthetic homogeneity (lots of white edges, touch of blues, red and green color, sans serif types). The first is technical; some of these platforms may be built using preexisting material, such as snippets of code used to streamline their construction. Secondly, designers often aim to make every new user's journey intuitive and to lower the entry threshold. Using the visual codes of known traditional social media platforms, such as Facebook, greatly assists with developing the capacity for effortless interaction. Though efficient, these variously traced design consistencies are, however, arguably somewhat counterintuitive—as the platforms they serve are and will likely always be used in diverse contexts, on diverse subjects, and emerge from diverse initiatives.

Nevertheless, these initially plain-looking interfaces, that seem to fit any context, present a readable assemblage of democracy's normative theories. Moreover, their relative consistency makes sense when considered in terms of a deliberative and participatory democracy approach, whereby deliberation can grant a form of legitimacy to processes of decision-making. However the omnipresence of voting functionalities they utilize connects them to survey tools and follows an aggregative logic, in which the majority of individual votes dictates the ultimate legitimacy of any given decision. Furthermore, the highlighting and emphasizing of opposition in these online assemblies conveys an agonistic conception of deliberation, which in turn illuminates the irreducible nature of conflict within democracy.

Analyzing online assemblies' design patterns helps us question ways of materializing specific conceptions of democracy such as aggregative, deliberative, and agonistic democracy. Furthermore, pursuing new pathways of analysis in line with these deliberations could help to seize design spaces not-yet seized or invested in by civic technologies, and open avenues for new forms of assemblies online.